\documentclass[12pt]{article}
\def\slash#1{\setbox0=\hbox{$#1$}#1\hskip-\wd0\hbox to\wd0{\hss\sl/\/\hss}}
\usepackage{epsf, cite, amsmath, amssymb}
\usepackage{epsfig}
\usepackage[all]{xy}
\makeatletter
\renewcommand\section{\@startsection {section}{1}{\z@}%
                                   {-3.5ex \@plus -1ex \@minus -.2ex}
                                   {2.3ex \@plus.2ex}%
                                   {\normalfont\large\bfseries}}
\renewcommand\subsection{\@startsection{subsection}{2}{\z@}%
                                     {-3.25ex\@plus -1ex \@minus -.2ex}%
                                     {1.5ex \@plus .2ex}%
                                     {\normalfont\bfseries}}
\makeatother

\let\non\nonumber

\newcommand{\bea}{\begin{eqnarray}}
\newcommand{\eea}{\end{eqnarray}}
\newcommand{\be}{\begin{equation}}
\newcommand{\ee}{\end{equation}}

\newcommand{\p}{\partial}

\newcommand{\s}{\sigma}

\newcommand{\C}[1]{$(\ref{#1})$}

\begin{document}

\begin{titlepage}

\begin{center}



\vskip 2 cm
{\Large \bf Higher Derivative Corrections in Holographic QCD}\\
\vskip 1.25 cm { Anirban Basu\footnote{email: abasu@ias.edu}
}\\
{\vskip 0.75cm
Institute for Advanced Study, Princeton, NJ 08540, USA\\
}

\end{center}

\vskip 2 cm

\begin{abstract}
\baselineskip=18pt

We consider the effect of the ${\cal{R}}^4$ term in type IIA string theory on the supergravity background
dual to $N_c$ D4 branes compactified on a circle with supersymmetry 
breaking boundary conditions.
We study the dynamics of D8 branes in this perturbed geometry in the probe approximation. 
This leads to an analysis of higher derivative corrections in holographic QCD beyond the
supergravity approximation. We make a rough estimate of the corrections to the masses 
of some of the
lightest (axial) vector mesons. The corrections are suppressed by a factor of 
$(g_{YM}^2 N_c)^{-3} $ compared to their supergravity values.
We find that the masses of these mesons increase from their supergravity values.

\end{abstract}

\end{titlepage}

\pagestyle{plain}
\baselineskip=18pt

\section{Introduction}

It is a challenging problem to understand strong coupling phenomena such as confinement and
chiral symmetry breaking in QCD at low energies. Using the gauge/string duality, a model of the
holographic dual of pure
QCD without matter was proposed in~\cite{Witten:1998zw}. 
Witten considered $N_c$ D4 branes with one of the 
world volume directions compactified on a circle, with anti--periodic boundary conditions for the fermions. 
This configuration breaks supersymmetry, and the fermions and the scalars on the world volume theory
of the D4 branes become massive at tree level and at one--loop level respectively. Thus at low energies, 
this reduces to a theory of pure Yang--Mills in four dimensions which is confining. Flavor
was added to this model in the probe approximation~\cite{Karch:2002sh}, by which one means that
$N_f$ flavor branes are placed in the background geometry dual to $N_c$ color branes, such 
that $N_f \ll N_c$. More recently,
Sakai and Sugimoto considered the dynamics of flavor D8 branes in the background geometry of color D4 branes
in the probe approximation~\cite{Sakai:2004cn,Sakai:2005yt}. 
This gives a model of holographic QCD with matter, 
which
differs from QCD at energies comparable to the Kaluza--Klein mass scale of the theory, which is
determined by the 
radius of the circle on which the D4 branes are compactified. Also the theory has an $SO(5)$ symmetry 
transverse to the color branes, unlike QCD. Nevertheless, this theory is an interesting model in trying
to understand QCD at energy scales below the Kaluza--Klein mass scale. 
Sakai and Sugimoto demonstrated chiral symmetry breaking in this theory by analyzing the
dynamics of the flavor branes in the background geometry of the color branes in the supergravity 
approximation. For related work, 
see~\cite{Son:2003et,Babington:2003vm,Kruczenski:2003uq,Aharony:2006da,Gepner:2006qy}. Letting the
radius of the circle on which the D4 branes are compactified to go to infinity, one obtains a theory 
with broken chiral symmetry, but which is unconfined. Such systems have been studied 
in~\cite{Antonyan:2006vw,Parnachev:2006dn,Gao:2006up,Antonyan:2006qy,Antonyan:2006pg,Basu:2006eb} 
(also see~\cite{Bak:2004nt} for a related discussion).   

Now understanding various aspects of holographic QCD has been so far done at the level of supergravity. In 
this paper, we attempt to go beyond the supergravity approximation, and include the effects of stringy 
corrections. In particular, we shall include the effect of the ${\cal{R}}^4$ term in the effective 
action of type IIA string theory. When considering only the supergravity contributions, we are working
in the large $N_c$ limit, with $g_{YM}^2 N_c \rightarrow \infty$, 
where $g_{YM}$ is the four dimensional gauge coupling. When we consider the 
contribution due to the ${\cal{R}}^4$ term, which is $\alpha'^3$ suppressed compared to
the supergravity contributions, we consider the leading correction to supergravity where
$g_{YM}^2 N_c$ is kept large but finite.   

We first consider the effect of the ${\cal{R}}^4$ term on the supergravity background 
dual to the D4 
branes. We analyze the perturbed geometry, and consider the dynamics of the D8 branes in this geometry,
still in the probe approximation. In particular, we focus on the dynamics of the gauge fields on the
world volume theory of the D8 branes. The fluctuations which are along the $(3+1)$ directions are massive,
and are interpreted as (axial) vector mesons~\cite{Sakai:2004cn} in holographic QCD. Using the perturbed 
geometry, we make a very rough estimate of the corrections to the masses of the lightest mesons, due to 
the higher derivative corrections. At the level of approximation we use, we find that the masses increase
from their supergravity values, with $\delta m^2 \sim (g_{YM}^2 N_c)^{-3}$. The analysis of the perturbed 
geometry is a generalization of the method in~\cite{Gubser:1998nz,Pawelczyk:1998pb} to non--conformal cases, 
where the effect of the ${\cal{R}}^4$ term on the near--horizon geometry dual to 
D3 branes in type IIB string theory
was considered (also see~\cite{Caldarelli:1999ar}). 
We shall see that the analysis gets considerably more complicated, essentially because 
of the loss of conformality.    
Though we obtain the exact perturbed geometry, we can solve for the coefficients in the various expressions
only recursively as we shall 
demonstrate below. Coupled with the fact that the mesons masses can only be calculated 
roughly at the supergravity 
level, the complexity of the equations allows us to obtain a rather rough estimate of the corrections to the 
meson masses.

It should be noted that our analysis does not give the complete answer due to the $\alpha'^3$ corrections to 
supergravity. We consider only the ${\cal{R}}^4$ term in the entire supermultiplet at $O(\alpha'^3)$, 
and there are many other
terms in this supermultiplet that will also contribute (for example, ${\cal{R}}^3 F_4^2$, ${\cal{R}}^3 (\p 
\phi)^2$, and so on\footnote{$F_4$ and $\phi$ are the R--R four form and the dilaton respectively.}). 
The entire ${\cal{R}}^4$ supermultiplet is not well understood (see~\cite{deHaro:2002vk,Policastro:2006vt} 
for a relevant discussion in type IIB string theory), and so we restrict 
ourselves to the ${\cal{R}}^4$ term only. Adding the other contributions, it is possible that the
values of various observables like meson masses will change.

\section{The background geometry dual to the color branes}

In order to study holographic QCD, we 
consider $N_c$ D4 branes extending along the directions $0,1,2$, and $3$, and compactified along the direction 
$x_4$, with antiperiodic boundary conditions for the fermions to break supersymmetry. The number of colors
$N_c$ is taken to be very large in the entire discussion.  

The dual supergravity background in the string frame is given 
by\footnote{This is obtained from the near extremal 
black 4---brane solution~\cite{Horowitz:1991cd} by 
interchanging the role of the time and $x_4$ coordinates.} 

\be \label{sugraact}
ds^2_{string} =   \Big( \frac{U}{R} \Big)^{3/2} \Big( \eta_{\mu\nu} dx^\mu dx^\nu
+ f (U) d x_4^2 \Big) + \Big( \frac{R}{U} \Big)^{3/2} \Big( \frac{dU^2}{f (U)} + U^2 d \Omega_4^2 
\Big) , \ee

where

\be \label{valR} e^\phi = g_s \Big( \frac{U}{R} \Big)^{3/4}, \quad R^3 = \pi g_s N_c \alpha'^{3/2}, 
\quad f (U) = 1-\frac{U_{KK}^3}{U^3},\ee

and $U$ is the radial coordinate transverse to the 4--brane world volume. 
The four form field strength is given by $F_4 = Q \omega_4$, where $\omega_4$ is 
the volume form on the unit four sphere. Using the Dirac quantization condition~\cite{Polchinski2},   

\be \label{Diracquant}
\int_{S^4} F_4 = \frac{\kappa_{10} N_c}{\sqrt{\pi \alpha'}}, \ee

we get that 

\be  \label{valQ}
Q = 3 N_c \pi \alpha'^{3/2} . \ee 

Including the ${\cal{R}}^4$ correction to the supergravity action, the relevant action in the string frame 
is given 
by~\cite{Green:1981yb,Green:1981ya,Gross:1986iv,D'Hoker:1988ta}\footnote{See
~\cite{Grisaru:1986vi,Freeman:1986zh,Park:1987jp}
for the analysis using conformal sigma model techniques.}

\be S_{string} = \frac{1}{2\kappa_{10}^2} \int d^{10}x \sqrt{-g} \Big[ e^{-2\phi} 
\Big( R + 4 (\nabla \phi)^2 
+ \gamma W \Big) - \frac{1}{2. 4!} F_4^2 \Big], \ee

where

\be \gamma = \frac{\zeta (3)}{8} \alpha'^3,\ee
and

\be W = C^{HMNK} C_{PMNQ} C_H^{~RSP} C^Q_{~RSK} + \frac{1}{2} C^{HKMN} C_{PQMN} C_H^{~RSP} C^Q_{~RSK},\ee

where $C^H_{~MNK}$ is the Weyl tensor. We now want to compute the perturbed background due to the
introduction of the ${\cal{R}}^4$ interaction. To do so, we find
it convenient to go to the Einstein frame which gives the action
 
\be \label{Einsact}
S_{Einstein} = \frac{1}{2\kappa^2} \int d^{10}x \sqrt{-\hat{g}} \Big[ \hat{R} -\frac{1}{2} 
(\hat\nabla \phi)^2 
- g_s^{3/2}
\frac{e^{\phi/2}}{2. 4!} {\hat{F}}_4^2 + g_s^{3/2} \gamma e^{-3\phi/2} \hat{W} \Big], \ee

where $\hat{g}$ is the Einstein frame metric, and $\kappa = \kappa _{10} g_s$. The metric 
${\hat{g}}_{MN}$ is given by

\be \label{metEin}
ds^2_{Einstein} = \Big( \frac{U}{R} \Big)^{9/8} \Big( \eta_{\mu\nu} dx^\mu dx^\nu
+ f (U) d x_4^2 \Big) + \Big( \frac{R}{U} \Big)^{15/8} \Big( \frac{dU^2}{f (U)} + U^2 d \Omega_4^2 
\Big). \ee

We make the ansatz for the perturbed metric 

\be \label{metricansatz}
ds^2_{Einstein} = H^2 (U) \Big[ K^2 (U) d x_4^2 + P^2 (U) d U^2 + \eta_{\mu\nu} dx^\mu dx^\nu \Big]
+ L^2 (U) d \Omega_4^2 .\ee

Thus translational invariance along the $(3+1)$ directions and the $x_4$ direction, as well as the 
transverse $SO(5)$ rotational invariance is preserved. We now want to construct the perturbed metric 
and the dilaton in the Einstein frame to leading order in $\gamma$. 
One might think that $F_4$ gets perturbed to $F_4 = Q
\lambda (U) \omega_4$, where $\lambda (U) = 1 + O(\gamma)$. However, the Dirac 
quantization condition \C{Diracquant} prevents this and $F_4$ remains unperturbed. 

Using the symmetries of the ansatz \C{metricansatz}, we see that the action \C{Einsact}
is given by

\bea \label{valmetric}
S_{Einstein} = V  \int_{U_{KK}}^\infty dU \sqrt{\tilde{\hat{g}}} \Big[ \hat{R} -\frac{1}{2} 
(\hat\nabla \phi)^2 - \frac{g_s^2}{2 \cdot 4!} \Big( \frac{U}{R} \Big)^{3/8} e^{(\phi_1 + \phi_2 + \ldots)/2}
{\hat{F}}_4^2 \non \\
+  \gamma \Big( \frac{U}{R} \Big)^{-9/8} e^{-3(\phi_1 + \phi_2 + \ldots)/2} 
\hat{W} + \Big], \eea

where

\be \sqrt{\tilde{\hat{g}}} = H^6 K P L^4,\ee

and

\be \label{valV} V = \frac{{\rm Vol(S^4)} V_{3,1} {\rm Vol(S^1)}}{2\kappa^2}.\ee

In \C{valV}, ${\rm Vol(S^4)} = 8\pi^2/3$ is the volume of the unit four sphere, and ${\rm Vol(S^1)}$ is the
circumference of the circle along $x_4$. Also we have expressed

\be \phi = \phi_0 + \frac{3}{4} {\rm ln} \Big( \frac{U}{R} \Big) + \phi_1 + \phi_2 + \ldots,\ee

where $g_s = e^{\phi_0}$, $\phi_1 \sim O(\gamma)$, $\phi_2 \sim O(\gamma^2)$, and the remaining terms 
in $\ldots$ are of $O(\gamma^3)$. Thus in \C{valmetric}, we have that

\be (\hat\nabla \phi)^2 = \frac{1}{H^2 P^2} \Big( \frac{9}{16U^2} + \frac{3}{2U} (\partial_U \phi_1) +
(\partial_U \phi_1)^2 + \frac{3}{2U} (\partial_U \phi_2) + O(\gamma^3) \Big). \ee

We describe the various details for obtaining the perturbed metric and the dilaton in the appendices.
Note that even though we do not have closed form expressions for the metric or the dilaton perturbations in
\C{valval} or \C{valdil}, we see that these perturbations take a simple form: they are given by sums of
positive integer powers of harmonic functions (upto a factor of $U^{-3/2}$). This is similar to what
happens in the conformal case of the 3--brane geometry~\cite{Gubser:1998nz,Pawelczyk:1998pb}. 

\section{The flavor branes in the background geometry}

Having obtained the perturbed background dual to the color D4 branes, we now
analyze the dynamics of $N_f$ probe D8 branes in this background geometry. We work in the probe
approximation so that $N_f \ll N_c$. Before taking the dynamics into account, the D brane configuration
is given by  

\begin{equation}
\nonumber
\begin{array}{cccccccccccccccccccccccccccccccccccccccc}
  &   &0& \ \  &1& \ \ &2& \ \ &3& \ \ &4& \ \  &5& \
\ &6& \ \ &7&  \ \ &8&  \ \ &9& \\
\\ 
 \text{D}4& :  &\textsf{x}& \  \ &\textsf{x}& \  \ &\textsf{x}& \
     \ &\textsf{x}&\  \ &\textsf{x}&  \\
 \\
 \text{D}8, &\overline{\text{D}8}  : &\textsf{x}& \ &\textsf{x}& \
&\textsf{x}& \ \ & \ \textsf{x} & \ \ & \  & \ \ & \ \textsf{x}\ & \ \ &\textsf{x}& \
\ &\textsf{x}& \ \ &\textsf{x}&  \ \ &\textsf{x}&  \ \  \\
\\
\end{array}
\label{intersection}
\end{equation}

Thus the flavor D8 and $\overline{{\rm{D}}8}$ branes intersect the color 
D4 branes along $(3 +1)$ dimensions and 
are separated along the $x_4$ direction. 
This configuration becomes very different when the dynamics are considered~\cite{Sakai:2004cn}.
The D8 and $\overline{{\rm{D}}8}$ branes get connected which is interpreted as chiral symmetry 
breaking. Our aim is to consider some aspects of the dynamics of the flavor branes in the perturbed 
background geometry dual to the color branes. First let us see how the flavor branes
deform in this background geometry. The background metric is given by

\bea \label{defmet}
d s^2_{string} &=&  \Big( \frac{U}{R} \Big)^{3/2} \Big[ e^{\gamma {\hat{\phi}}_1/2}
\eta_{\mu\nu} dx^\mu dx^\nu
+ f (U)  e^{\gamma( {\hat{\phi}}_1/2 + 2 a_1 + 18 b_1)} d x_4^2 \Big] \non \\
&&+ \Big( \frac{R}{U} \Big)^{3/2} \Big[ 
e^{\gamma ( {\hat{\phi}}_1/2 + 2 b_1)} \frac{dU^2}{f (U)} + U^2 e^{ \gamma ({\hat{\phi}}_1/2 
+ 2 c_1)}  d \Omega_4^2 \Big] .\eea

Considering probe D8 branes in this geometry where 
$U = U (x_4)$, we see that the induced metric on the D8 brane world volume is given by

\bea 
\label{indmet}
d s^2_{D8} &=&  \Big[ \Big( \frac{U}{R} \Big)^{3/2}
f (U) e^{\gamma( {\hat{\phi}}_1/2 + 2 a_1 + 18 b_1)}  +  \Big( \frac{R}{U} \Big)^{3/2}
e^{\gamma ( {\hat{\phi}}_1/2 + 2 b_1)}  \frac{U'^2}{f (U)} \Big] d x_4^2 \non \\
&& +\Big( \frac{U}{R} \Big)^{3/2} e^{\gamma {\hat{\phi}}_1/2} 
\eta_{\mu\nu} dx^\mu dx^\nu + \Big( \frac{R}{U} \Big)^{3/2}  U^2 e^{\gamma ({\hat{\phi}}_1/2 
+ 2 c_1)}  d \Omega_4^2 , \eea

where $U' = (\p U/\p x_4)$. In the expressions \C{defmet} and \C{indmet} and in the ones that follow,
the exponentials are only to be expanded to $O(\gamma)$. 
Inserting the induced metric \C{indmet} into the DBI action of the D8 branes, we get that

\be \label{DBIact}
S_{D8} \sim \int d^4 x d x_4 e^{2\gamma ({\hat{\phi}}_1/2 
+ 2 c_1) (U)} U^4 \sqrt{ e^{\gamma( {\hat{\phi}}_1/2 + 2 a_1 + 18 b_1)(U)} f (U) + \Big( \frac{R}{U} \Big)^3
e^{\gamma ( {\hat{\phi}}_1/2 + 2 b_1)(U)}  \frac{U'^2}{f (U)}}.\ee

Following Sakai and Sugimoto, we look for a solution to the classical equation of motion for the 
D8 brane resulting from \C{DBIact} which asymptotes as $U \rightarrow 
\infty$ to a fixed value of $x_4$. This is the position of the flavor brane along the $x_4$ direction in 
the naive picture. To consider a configuration of D8 and $\overline{{\rm{D}}8}$ branes which connect together
leading to chiral symmetry breaking, we also want this configuration to satisfy 
$U(x_4 = 0) = U_0 ,U'(x_4 = 0) = 0$, such that it is symmetric about $x_4 =0$. Thus naively,
without considering the dynamics, this corresponds to D8 and $\overline{{\rm{D}}8}$ branes
placed symmetrically about $x_4 =0$. Including the dynamics, they get connected.
This solution which has the interpretation of a wormhole solution connecting the D8 and 
$\overline{{\rm{D}}8}$ branes asymptotically exhibits chiral symmetry breaking in holographic QCD. 
The throat of this wormhole has its minimum radius $U_0$ at $x_4 =0$. This solution is given by

\be \label{shapebrane}
x_4 (U) = U_0^4 \sqrt{\frac{f (U_0)}{\Theta_2 (U_0)}} \int_{U_0}^U dU \frac{e^{\gamma ( {\hat{\phi}}_1/2 + 
2 b_1)(U)/2}}{f(U) \Big( \frac{U}{R} \Big)^{3/2} \sqrt{U^8 f(U) \frac{\Theta_1 (U)}{\Theta_1 (U_0))} 
- U_0^8 f(U_0) \frac{\Theta_2 (U)}{\Theta_2 (U_0)}}}, \ee

where 

\be \Theta_1 (U)= e^{2\gamma(3{\hat{\phi}}_1/2 + 2 a_1 + 18 b_1 + 4 c_1) (U)}, \quad 
\Theta_2 (U)= e^{\gamma({\hat{\phi}}_1/2 + 2 a_1 + 18 b_1) (U)}.\ee

Thus chiral symmetry continues to be broken in the presence of the higher derivative corrections.
In \C{shapebrane}, $U_0$ is an arbitrary parameter satisfying $U_0 \geq U_{KK}$. For the sake of simplicity, we 
shall analyze the spectrum for the case $U_0 = U_{KK}$ as done by Sakai and Sugimoto. Taking only the supergravity
action into account, they showed that the D8 and $\overline{{\rm{D}}8}$ branes are placed at antipodal points 
on the circle parametrized 
by $x_4$ when $U_0 = U_{KK}$. Now even when the higher derivative corrections are turned on, the  
ansatz we have made for the various perturbations preserves the same symmetries, in particular, translational 
invariance along the $x_4$ circle is preserved. 
Thus when $U_0 = U_{KK}$, we expect the branes to remain at antipodal
points on the $x_4$ circle.  We now show this is the case.  
 
Given the perturbed metric \C{defmet}, in order to avoid a conical singularity at 
$U = U_{KK}$, $x_4$ must be a periodic
variable satisfying

\be x_4 \sim x_4 + \frac{4 \pi R^{3/2} e^{-\gamma(a_1 + 8 b_1) (U_{KK})}}{3 \sqrt{U_{KK}}}. \ee

Thus the circumference of the $x_4$ circle is given by

\be \delta x_4 = \frac{4 \pi R^{3/2} e^{-\gamma(a_1 + 8 b_1) (U_{KK})}}{3 \sqrt{U_{KK}}}. \ee 

From \C{shapebrane}, we can calculate the position in $x_4$ where the D8 brane is placed which is given by 
$x_4 (\infty)$, and the D8 and $\overline{{\rm{D}}8}$ branes are separated by twice this 
distance. However, for $U_0 = U_{KK}$,
$f (U_0) = 0$, and the integrand in \C{shapebrane} diverges at $U = U_0$. Thus the expression for $x_4 (\infty)$
needs to be regularized. We regularize it by setting $\s = (U_0 / U_{KK}) = 1 + \epsilon$, and taking the limit 
$\epsilon \rightarrow 0$. We need to pick out the $O(1/\sqrt{\epsilon})$ terms from the integral which cancel the 
$O(\sqrt{\epsilon})$ term from $\sqrt{f(U_0)}$. We have that

\be 
x_4 (\infty) \vert_{U_0 = U_{KK}} 
= \frac{R^{3/2}}{\sqrt{U_{KK}}} \sqrt{\frac{g (\s)}{\Theta_2 (\s)}} \int_{\s = 1 + \epsilon}^\infty 
du \frac{e^{\gamma ( {\hat{\phi}}_1/2 
+ 2 b_1)(u)/2}}{g(u) u^{3/2} \sqrt{u^8 g(u) \frac{\Theta_1 (u)}{\Theta_1 (\s)} 
- \s^8 g(\s) \frac{\Theta_2 (u)}{\Theta_2 (\s)}}}, \ee
 
where 

\be g (u) = 1 - \frac{1}{u^3} . \ee

It is straightforward to show that the relevant contribution is contained in the expression given by

\bea 
x_4 (\infty) \vert_{U_0 = U_{KK}} 
&=& \frac{R^{3/2}}{3 \sqrt{U_{KK}}} \sqrt{\frac{3 \epsilon}{\Theta_2 (1)}} \int_{3\epsilon}^1 
dx \frac{e^{\gamma ( {\hat{\phi}}_1/2 
+ 2 b_1)(1)/2}}{x^{3/2}} \Big(1 - \frac{3 \epsilon}{x}\Big)^{-1/2} \non \\
&=& \frac{R^{3/2}}{3 \sqrt{U_{KK}}} 
\sqrt{\frac{ e^{\gamma ( {\hat{\phi}}_1/2 
+ 2 b_1)(1) }}{\Theta_2 (1)}} \sum_{k= 0}^\infty \frac{\Gamma (k + 1/2)}{\Gamma (1/2) k ! (k + 1/2)} \non \\
&=&  \frac{\pi R^{3/2} e^{-\gamma(a_1 + 8 b_1) (U_{KK})}}{3 \sqrt{U_{KK}}} = \frac{\delta x_4}{4}, \eea
 
where by equality in the first two lines we mean that we only keep the relevant 
terms which are non--vanishing in the limit $\epsilon \rightarrow 
0$. Thus on adding the higher derivative corrections, the radius of $x_4$ changes, but the D8 and $\overline{D8}$ 
branes continue to be at antipodal points on the $x_4$ circle. 

\section{Estimating the corrections to (axial) vector meson masses}

Having discussed the flavor brane configuration, we now consider the effect of the higher derivative 
corrections on the masses of the (axial) vector mesons in holographic QCD. These vector mesons are 
obtained from the fluctuations of the gauge fields on the D8 brane world volume
in the background geometry. In fact, the fluctuations along the directions $x_\mu$ yield the vector
mesons, while the fluctuations along $U$ yield the pions. 
Expanding the DBI action to quadratic order in the gauge fields, we get that

\be \label{gaugeact}
S_{D8} =  - \frac{T_8}{4} \int d^4 x d U e^{\gamma ({\hat{\phi}}_1/4 + b_1 + 4 c_1) (U)} {\rm Tr} \Big[ 
\frac{R^{9/2}}{\sqrt{U f (U)}} {\hat{F}}_{\mu\nu}^2 + 2e^{-2\gamma b_1 (U)} U^{5/2} R^{3/2} \sqrt{f 
(U)} {\hat{F}}^\mu~_U F_{\mu U} \Big], \ee

where
 
\be T_8 = \frac{{\rm Vol} (S^4)}{(2 \pi)^6 \alpha'^{5/2} g_s}, \ee 

and $\hat{F}$ means that the four dimensional indices are raised with the Minkowski metric. We focus only on the 
vector mesons and drop the terms involving the pions. Thus defining

\be A_\mu (x, U) = \sum_{n=1}^\infty B_\mu^{(n)} (x) \psi_n (U), \ee

we get that

\bea 
F_{\mu \nu} (x, U) = \sum_{n=1}^\infty F_{\mu\nu}^{(n)} (x) \psi_n (U), \non \\
F_{\mu U} (x, U) = -\sum_{n=1}^\infty B_\mu^{(n)} (x) \p_U \psi_n (U),
\eea

where $F_{\mu\nu}^{(n)} = \p_\mu B_\nu^{(n)} - \p_\nu B_\mu^{(n)}$. Thus we get that

\bea \label{pertact}
 S_{D8} &=&  - \frac{T_8}{4} \int d^4 x d U e^{\gamma ({\hat{\phi}}_1/4 + b_1 + 4 c_1) (U)} 
\sum_{m,n =1}^\infty {\rm Tr} \Big[ 
\frac{R^{9/2}}{\sqrt{U f (U)}} {\hat{F}}^{\mu\nu (m)} (x) F_{\mu\nu}^{(n)} (x) \psi_m (U) \psi_n  (U) \non \\&&
+ 2 e^{-2\gamma b_1 (U)} U^{5/2} R^{3/2} \sqrt{f (U)} {\hat{B}}^{\mu (m)} (x) B_\mu^{(n)} (x) (\p_U \psi_m) (U) 
(\p_U \psi_n) (U) \Big].
\eea

Note that \C{pertact} is independent of $a_1$. Defining 

\be Z^2 = \frac{U^3}{U_{KK}^3} -1, \quad K (Z) = 1 + Z^2, \ee
 
we get that

\bea \label{pertact2}
 S_{D8} &=&  - {\hat{T}}_8 \int d^4 x d Z e^{\gamma ({\hat{\phi}}_1/4 + b_1 + 4 c_1) (Z)} 
\sum_{m,n =1}^\infty {\rm Tr} \Big[\frac{1}{4} K^{-1/3} (Z) {\hat{F}}^{\mu\nu (m)} (x) F_{\mu\nu}^{(n)} (x) 
\psi_m (Z) \psi_n  (Z) \non \\&&
+ \frac{9 U_{KK}}{8 R^3} e^{-2\gamma b_1 (Z)}  K (Z) {\hat{B}}^{\mu (m)} (x) B_\mu^{(n)} (x) (\p_Z \psi_m) (Z) 
(\p_Z \psi_n) (Z) \Big],
\eea

where

\be {\hat{T}}_8 = \frac{2 R^{9/2} \sqrt{U_{KK}} {\rm Vol} (S^4)}{3 (2 \pi)^6 \alpha'^{5/2} g_s}. \ee

Canonical normalization of the kinetic term in \C{pertact2} leads to

\be \label{normkin}
{\hat{T}}_8 \int d Z e^{\gamma ({\hat{\phi}}_1/4 + b_1 + 4 c_1) (Z)}  K^{-1/3} (Z) 
\psi_m (Z) \psi_n  (Z) = \delta_{m,n}, \ee

while canonical normalization of the mass term gives

\be \label{normmass}
{\hat{T}}_8 \int d Z e^{\gamma ({\hat{\phi}}_1/4 - b_1 + 4 c_1) (Z)}  K (Z) 
(\p_Z \psi_m) (Z) (\p_Z \psi_n)  (Z) = \lambda_n \delta_{m,n}, \ee

where the mass is given by $m_n^2= (9 U_{KK} \lambda_n)/(4 R^3)$. Thus \C{pertact2} reduces to

\be
 S_{D8} =  -  \int d^4 x  
\sum_{m =1}^\infty {\rm Tr} \Big[\frac{1}{4} {\hat{F}}^{\mu\nu (n)}  F_{\mu\nu}^{(n)} 
+ \frac{m_n^2}{2}  {\hat{B}}^{\mu (n)} B_\mu^{(n)} \Big].
\ee

From \C{normkin} and \C{normmass} we obtain the eigenvalue equation for $\psi_n$ 

\be \label{eigeneqn}
e^{-\gamma ({\hat{\phi}}_1/4 + b_1 + 4 c_1)} K^{1/3}  \p_Z \Big[ 
e^{\gamma ({\hat{\phi}}_1/4 - b_1 + 4 c_1) } K (\p_Z \psi_n )\Big] = - \lambda_n
\psi_n . \ee

Now because the perturbations ${\hat{\phi}}_1 ,b_1$ and $c_1$ are functions of $Z^2$, note that the 
action \C{pertact2} is invariant under $(x^\mu, Z) \rightarrow (-x^\mu, -Z)$, which has the interpretation of 
space time parity~\cite{Sakai:2004cn}. 
Thus from \C{eigeneqn} we see that $\psi_n$ has definite parity under $Z \rightarrow -Z$.  
The aim is to construct $\psi_{n}$ which has even (odd) parity for $n$ odd (even).  Thus $B_\mu^{(n)}$ is a vector 
(axial vector) if $n$ is odd (even). Under charge conjugation, $B_\mu^{(n)}$ is even (odd) if $n$ is even
(odd)~\cite{Sakai:2004cn}. We shall 
focus on the lowest lying modes with $n=1,2$ which have $(C,P)= (-,-)$ and $(+,+)$ respectively. 
In holographic QCD, these modes are to be identified with the 
$\rho$ meson and the $a_1 (1260)$ meson respectively. 

In order to find the correction to the vector meson masses due to the higher derivative corrections,
we write 

\bea \psi_n = \psi_n^{(0)} + \frac{\gamma}{\sqrt{R^9 U_{KK}^3}} 
\psi_n^{(1)}, \non \\ 
\lambda_n = \lambda_n^{(0)} + \frac{\gamma}{\sqrt{R^9 U_{KK}^3}} 
\lambda_n^{(1)}, \eea 

where $\psi_n^{(0)}$ and $\lambda_n^{(0)}$ are the supergravity expressions, while
$\psi^{(1)}_n$ and $\lambda_n^{(1)}$ are the higher derivative corrections. 

Using first order perturbation theory, \C{normkin}, and \C{eigeneqn}, we get that

\be \label{massform}
\lambda_n^{(1)}  = - {\hat{T}}_8 \int d Z \psi_n^{(0)} (Z) {\hat{H}}_n (Z) \psi_n^{(0)} (Z) ,\ee

where

\bea {\hat{H}}_n (Z) &=& K \Big( \frac{{\hat{\phi}}_1'}{4} - b_1' + 4 c_1' \Big) 
\frac{\p}{\p Z} - 2 b_1 \Big( 2 Z \frac{\p}{\p Z} + K \frac{\p^2}{\p Z^2}\Big)\non \\
&=&   K \Big( \frac{{\hat{\phi}}_1'}{4} - b_1' + 4 c_1' \Big) 
\frac{\p}{\p Z} + 2 \lambda_n^{(0)} b_1 K^{-1/3}, \eea

where we have used the $O(1)$ relation from \C{eigeneqn}

\be \label{leadeqn}
K^{1/3} ( K \p _Z^2 \psi_n^{(0)} + 2 Z \p_Z \psi_n^{(0)} ) = - \lambda_n^{(0)} \psi_n^{(0)} .\ee

Integrating \C{massform} by parts, one can also express $\lambda_n^{(1)}$ as

\be \label{massform2}
\lambda_n^{(1)}  =  {\hat{T}}_8 \int d Z K \Big( \frac{{\hat{\phi}}_1}{4} - b_1 + 4 c_1 \Big) 
(\p_Z \psi_n^{(0)})^2 - \lambda_n^{(0)} {\hat{T}}_8 \int d Z K^{-1/3} 
\Big( \frac{{\hat{\phi}}_1}{4} + b_1 + 4 c_1 \Big) (\psi_n^{(0)})^2 .\ee

In \C{massform} and the equations that follow, we have removed an overall factor 
of $1/{\sqrt{R^9 U_{KK}^3}}$ from ${\hat{\phi}}_1, b_1$ and $c_1$ for notational simplicity.
Now it is difficult to calculate $\lambda_n^{(1)}$ from \C{massform} (or \C{massform2})
exactly because the unperturbed wavefunction
$\psi_n^{(0)} (Z)$ is not known exactly. Also although we have recursion relations
for the coefficients describing the perturbed geometry, we do not have closed form 
expressions for them. 
So we will make a very rough estimate of the correction to the meson masses, which we turn to now.

Sakai and Sugimoto considered normalizable wavefunctions satisfying the Schrodinger equation
\C{leadeqn}, and the normalization condition \C{normmass}, and 
obtained $\lambda_n^{(0)}$ numerically using the shooting technique. They obtained

\be \lambda_1^{(0)} \approx 0.67^{(-,-)}, \qquad \lambda_2^{(0)} \approx 1.6^{(+,+)} ,\ee

for the two lightest modes that satisfy \C{leadeqn}. Now for low values of $n$, these normalizable
wave functions must be concentrated around $Z =0$, while they spread out more and more to larger values of 
$Z$ as $n$ increases. 
Since we will focus on the two lowest lying normalizable eigenstates 
of \C{leadeqn}, the correction to the meson masses given by \C{massform} (or \C{massform2}) should receive 
the maximum contribution from the neighbourhood of $Z = 0$ in the integral. Thus in order to make
a rough estimate of $\lambda_n^{(1)}$,
we shall focus on this region only. Of course, as larger and larger values of $Z$ are considered, the 
approximation gets better and better. 

So in the various expressions, we focus on the region of integration around $Z=0$. In fact,  
we shall make the crudest approximation, and restrict ourselves to terms only upto 
$O(Z)$ in the various expressions. Now it is easy to construct 
an approximate normalized wavefunction which solves \C{leadeqn} at small $Z$.
Keeping terms only upto $O(Z)$, \C{leadeqn} reduces to

\be \p _Z^2 \chi_n (Z) \approx ( 1- \lambda_n^{(0)} ) \chi_n (Z), \ee

where $ \psi_n^{(0)} (Z) = e^{-Z^2/2} \chi_n (Z)$. Thus the approximate normalized
wave functions for the two lowest modes are given by

\be \label{psione}
\psi_1^{(0)} (Z) \approx \frac{0.69}{\sqrt{{\hat{T}}_8}} e^{-Z^2/2} {\rm cosh} \Big( 
\frac{Z}{\sqrt{3}} \Big), \ee

and 

\be \label{psitwo} \psi_2^{(0)}  (Z) \approx \frac{1.58}{\sqrt{{\hat{T}}_8}} e^{-Z^2/2} {\rm sin} \Big( 
\sqrt{\frac{3}{5}} Z\Big). \ee

In order to estimate $\lambda_n^{(1)}$ at this order, we also need approximate 
expressions for ${\hat{\phi}}_1 (Z), b_1 (Z)$ and $c_1 (Z)$. Because ${\hat{\phi}}_1 
(Z), b_1 (Z)$ and $c_1 (Z)$ are functions of $Z^2$ only, and we are restricting to terms of $O(Z)$,
we can replace ${\hat{\phi}}_1 (Z) , b_1 (Z)$ 
and $c_1 (Z)$ by the constant modes ${\hat{\phi}}^\star , b^\star$ and $c^\star$ respectively, 
where\footnote{In fact, we can look at the system of equations
\C{moreeqnabc} at small $Z$, keeping terms only to $O(Z)$. Noting
from the exact expressions for the perturbations in terms of $\eta$ that their behavior is similar at small
$Z$, one can solve them directly, and obtain the solutions for ${\hat{\phi}}_1 (Z), b_1 (Z)$ and $c_1 (Z)$
as described above. Note that the $Z$ dependence of the perturbations obtained by solving \C{moreeqnabc} 
at small $Z$, is very different from that in the wave functions \C{psione} and \C{psitwo} obtained by solving
\C{leadeqn} at small $Z$.}

\be {\hat{\phi}}_1 (Z) = {\hat{\phi}}^\star + O(Z^2), \quad b_1 (Z) = b^\star + O(Z^2), \quad
\quad c_1 (Z) = c^\star + O(Z^2). \ee

Thus using \C{normkin} and \C{normmass}, from \C{massform} (or \C{massform2}), we get that

\be \label{valapprox}
\lambda_n^{(1)} \approx - 2 \lambda_n^{(0)} b^\star. \ee

Thus at this order

\be \frac{\delta m_2^2}{\delta m_1^2} \approx  \frac{\lambda_2^{(0)}}{\lambda_1^{(0)}} \approx 2.4 .
\ee

To obtain the masses,
we estimate $b^\star$ using \C{moreeqnabc} directly. Setting $a_1 (Z) = a^\star , b_1 (Z) = 
b^\star$ and $c_1 (Z) = c^\star$ and equating terms of $O(1)$, we get that\footnote{Of the 
three resulting equations from \C{moreeqnabc}, only two are linearly independent: the first two equations 
are the same.}

\be \label{approxval}
b^\star \approx -20.6 , \quad c^\star \approx -9.2 .\ee 

Note that the zero mode $a^\star$ is undetermined by the equations. 

In order to get better estimates, one has to keep terms at higher orders in $Z$, and solve for the wavefunction, as 
well as the metric and dilaton perturbations using  \C{moreeqnabc} and \C{leadeqn}. 
The $Z$ dependence of these quantities is going to be different, and 
the values of the various coefficients (for example, the constant terms in $b_1$ and $c_1$) are going to change 
too. However, on including the various contributions, $\lambda_n^{(1)}$ should not 
change by a large amount as the low lying states are localized 
around $Z =0$. It would be interesting to include higher powers of $Z$ in this analysis, and try to get a 
better estimate. Presumably keeping a reasonably small number of terms in the expansion in $Z$ will 
make the estimates converge to a sharp value of $\lambda_n^{(1)}$.   
Using our rough estimates, we get that
\bea
\frac{\delta m_1^2}{M_{KK}^2} \approx 112 (g_{YM}^2 N_c)^{-3}, \non \\
\frac{\delta m_2^2}{M_{KK}^2} \approx 267 (g_{YM}^2 N_c)^{-3},
\eea    

where $M_{KK} \equiv 2\pi/ \delta x_4\vert_{(\gamma =0)} = (3 \sqrt{U_{KK}})/(2 R^{3/2})$, and we 
have used the relations~\cite{Kruczenski:2003uq}

\be R^3 = \frac{g_{YM}^2 N_c \alpha'}{2 M_{KK}} , \quad  U_{KK} = \frac{2g_{YM}^2 N_c M_{KK}
\alpha'}{9}. \ee

Thus at the level of the approximations we have made, we see that the masses of the two lightest (axial)
vector mesons increase from their supergravity values. In fact, using gauge/gravity duality techniques
to calculate heavy quark potentials, one obtains that $g_{YM}^2 N_c \approx 1$ 
(see~\cite{Andreev:2006ct}, for example).
Thus our rough estimates give us that
\bea
\frac{\delta m_1^2}{M_{KK}^2} \approx 112 , \quad
\frac{\delta m_2^2}{M_{KK}^2} \approx 267 .
\eea 
 
Trying to analyze the effect of the higher derivative corrections in other applications of holographic QCD
is an important problem in general. It would be nice to have exact expressions for the metric and dilaton
perturbations, as that will make calculations more concrete and predictive. In order to make precise
quantitative predictions in holographic QCD due to corrections to supergravity at $O(\alpha'^3)$, it 
is also important to understand the detailed structure of the ${\cal{R}}^4$ supermultiplet in type IIA string 
theory.  

\section*{Acknowledgements}

I would like to thank I.~Klebanov, J.~Maldacena, A.~Maloney, D.~Mateos, and S.~Sugimoto for useful comments. I am 
particularly thankful to A.~Maharana for many useful discussions, and for technical assistance in computing 
$\hat{W}$.
The work of A.~B. is supported by NSF Grant No.~PHY-0503584 and the William D. Loughlin membership.

\section{Appendix}

In this appendix, we describe the construction of the perturbed metric and the dilaton due to the
higher derivative corrections.

\appendix

\section{Obtaining the perturbed metric in the Einstein frame}

The aim is to first solve \C{valmetric} to leading order in $\gamma$ and find the 
perturbed metric in the Einstein frame. Considering \C{valmetric} at $O(\gamma)$, we get that

\be S_{Einstein}^{O(\gamma)} = S_0^\gamma + S_1^\gamma ,\ee

where $S_1^\gamma$ contains $\phi_1$, and $S_0^\gamma$ is independent of it. Thus

\be \label{metpert}
S_0^\gamma = V  \int_{U_{KK}}^\infty dU \sqrt{\tilde{\hat{g}}} \Big[ \hat{R} -\frac{9}{32U^2 H^2 P^2} 
-\frac{g_s^2 Q^2}{2L^8} \Big( \frac{U}{R} \Big)^{3/8} +\gamma \Big( \frac{U}{R} \Big)^{-9/8} \hat{W} 
\Big],\ee

and

\be \label{phione}
S_1^\gamma = V  \int_{U_{KK}}^\infty dU \sqrt{\tilde{\hat{g}}} \Big[ -\frac{3\p_U
\phi_1}{4U H^2 P^2} - \frac{g_s^2 Q^2}{4 L^8} \Big( \frac{U}{R} \Big)^{3/8} \phi_1 \Big]. \ee

Since $\phi_1 \sim O(\gamma)$, we can replace the other fields in \C{phione} by their supergravity values.
Thus we get

\be 
S_1^\gamma= V \int_{U_{KK}}^\infty dU \Big[ -\frac{3 f(U) U^3\p_U \phi_1}{4} - \frac{g_s^2 Q^2 U^2 
\phi_1}{4 R^6} \Big],\ee

which vanishes on integrating by parts the first term, and using \C{valR} and \C{valQ}. So in order
to find the metric perturbation to $O(\gamma)$, we only need to consider \C{metpert}. Note that the 
dilaton perturbation to leading order is undetermined at $O(\gamma)$ in the perturbative expansion. 

To evaluate \C{metpert}, we use the parametrizations

\be \label{param}
H (U) = \Big( \frac{U}{R} \Big)^{9/16}, \quad K (U) = e^{a (U)+ \lambda b (U)}, \quad
P (U) = e^{b (U)}, \quad L (U) = e^{c (U)} U \Big( \frac{R}{U} \Big)^{15/16}, \ee

where $\lambda$ is a constant, which we now fix to simplify calculations. So in \C{param}, 
we have $c (U) \sim O (\gamma)$.
Using \C{param}, we calculate 
$\sqrt{\tilde{\hat{g}}} \hat{R}$ and get 

\bea \sqrt{\tilde{\hat{g}}} \hat{R} &=& -\frac{e^{a + (\lambda -1)b + 4c} \sqrt{UR}}{8}
\Big[ \frac{21 R}{4} -96 U^2 R (c')^2 -96 e^{2(b-c)} U \Big( \frac{U}{R} \Big)^{2} 
\non \\&&+5 UR a'
+ 5 (\lambda -9) UR b' -12 UR c' -64 U^2 R c' (a' +\lambda b' )   \Big] \non \\&&
- 2 R^{3/2} \frac{d}{dU} \Big[ U^{5/2} \Big( a' + \lambda b' + 4c' + \frac{1}{4U}\Big) 
e^{a + (\lambda -1)b + 4c}\Big].\eea 

Note that for $\lambda =9$, the coefficient of the $URb'$ term vanishes, and this is the value we choose.
Thus \C{metpert} yields

\be \label{solveact}
S_0^\gamma = V  \int_{U_{KK}}^\infty dU \Big[ l (a,a',b,b',c,c') +\gamma w (a,a',a'',b,b',b'',c,c',c'') 
\Big],\ee

where

\bea l (a,a',b,b',c,c') &=& -\frac{e^{a + 8b + 4c} \sqrt{UR}}{8}
\Big[ \frac{21 R}{4} -96 U^2 R (c')^2 -96 e^{2(b-c)} U \Big( \frac{U}{R} \Big)^{2} 
\non \\ &&+5 UR a' -12 UR c' -64 U^2 R c' (a' +9 b' )   \Big] \non \\&&
-\frac{9}{32} U^2 e^{a +8b+4c} \Big[ \Big( \frac{R}{U} \Big)^{3/2}
+  16 \Big( \frac{U}{R} \Big)^{3/2} e^{2b-8c} \Big],\eea

where we have dropped the total derivative, and 

\be w (a,a',a'',b,b',b'',c,c',c'') = U (U R)^{3/2} e^{a + 10b + 4c} 
\hat{W} (a,a',a'',b,b',b'',c,c',c'') .\ee

So we need to solve the Euler--Lagrange equations of motion arising from \C{solveact}
which are given by

\be \frac{\p l}{\p \xi_i} - \frac{d}{d U} \Big( \frac{\p l}{\p \xi_i'}\Big) = -\gamma \Big[ 
\frac{\p w}{\p \xi_i} -\frac{d}{d U} \Big( \frac{\p w}{\p \xi_i'}\Big) + \frac{d^2}{d U^2} 
\Big( \frac{\p w}{\p \xi_i''}\Big)\Big],\ee

where $\xi_i = a,b,c$. The equations of motion for $a, b$ and $c$ are given by

\bea \label{eqnabc}
\Big( \frac{U}{R} \Big)^{3/2} \Big[ 3 e^{2b} U^2 (8 e^{6c} -3) - 2 R^3 e^{8c} \Big(  8U c''
-b' (5 + 8Uc') + 4c' (4 + 5Uc') \Big) \Big] \non \\
= -\frac{2\gamma}{e^{a+8b-4c}} \Big[ 
\frac{\p w}{\p a} -\frac{d}{d U} \Big( \frac{\p w}{\p a'}\Big) + \frac{d^2}{d U^2} 
\Big( \frac{\p w}{\p a''}\Big)\Big], \non \\
\sqrt{\frac{U}{R}} \Big[ R^2 e^{8c} \Big( 15 R + 2UR a' (5 + 8Uc') + 48 URc'(7 + 8Uc') + 144 U^2 R c'' 
\Big) \non \\
+10 e^{2b} \Big( 9 - 24 e^{6c} \Big) U^3 \Big] =
 \frac{2 R \gamma}{e^{a+8b-4c}} \Big[ 
\frac{\p w}{\p b} -\frac{d}{d U} \Big( \frac{\p w}{\p b'}\Big) + \frac{d^2}{d U^2} 
\Big( \frac{\p w}{\p b''}\Big)\Big], \non \\
\sqrt{\frac{U}{R}}  \Big[ 6 U^3 e^{2b} (3 + 4 e^{6c}) - 2R^3 e^{8c} \Big\{ 
3 + 2 U\Big( 2U a'^2 + 15 c' + 2 a' (3 + 17 U b' + 3 U c') \non \\ + 2
[ 72 U b'^2 + 24 b' (1 + Uc') + U(6 c'^2 + a'' + 9 b'' + 3c'') ]\Big) \Big\} \Big] 
\non \\= -\frac{ R \gamma}{e^{a+8b -4c}} \Big[ 
\frac{\p w}{\p c} -\frac{d}{d U} \Big( \frac{\p w}{\p c'}\Big) + \frac{d^2}{d U^2} 
\Big( \frac{\p w}{\p c''}\Big)\Big], \eea

respectively. 
We now expand \C{eqnabc} to $O(\gamma)$. Since the right hand side
is already of $O(\gamma)$, we simply substitute the values in the supergravity solution. 
Defining

\be  \frac{U_{KK}}{U} \equiv \eta ,\ee

the relevant expressions are

\bea && \frac{\p w}{\p a} - \frac{d}{d U} \Big( \frac{\p w}{\p a'}\Big) + \frac{d^2}{d U^2} \Big( 
\frac{\p w}{\p a''}\Big) \non \\
&=& \frac{3 \sqrt{U R}}{512 R^5} \Big[ \frac{729}{2} - 2106 \eta^3  
+ \frac{22221}{2} \eta^6 -474174 \eta^9 + \frac{1072869}{2} \eta^{12} \Big], \non \\
&& \frac{\p w}{\p b} - \frac{d}{d U} \Big( \frac{\p w}{\p b'}\Big) + \frac{d^2}{d U^2} \Big( 
\frac{\p w}{\p b''}\Big) \non \\
&=& \frac{3 \sqrt{U R}}{512 R^5} \Big[ 3753  - 19008 \eta^3  
+ 105831 \eta^6 -4273602 \eta^9 + 4899321 \eta^{12} \Big], \non \\
&& \frac{\p w}{\p c} - \frac{d}{d U} \Big( \frac{\p w}{\p c'}\Big) + \frac{d^2}{d U^2} \Big( 
\frac{\p w}{\p c''}\Big) \non \\
&=& -\frac{3 \sqrt{U R}}{256 R^5} \Big[ 945  + 216 \eta^3  
- 20106 \eta^6 + 37581 \eta^9 - 16569 \eta^{12} \Big]. \eea

On the left
hand side of \C{eqnabc}, the contributions due to terms of $O(1)$ vanish, and the $O(\gamma)$ terms are the
leading effects. Defining the order $\gamma$ perturbations to the metric by

\bea a (U) &=& -\frac{27}{2} {\rm {ln}} \Big( \frac{R}{U} \Big) + 5 {\rm {ln}} f (U) +
\gamma a_1 (U), \non \\ b (U) &=& \frac{3}{2} {\rm {ln}} \Big( \frac{R}{U} \Big) -\frac{1}{2} 
{\rm {ln}} f (U)+ \gamma b_1 (U), \non \\ c (U) &=& \gamma c_1 (U),\eea

where $a_1 , b_1,$ and $c_1$ are $O(1)$, substituting them into \C{eqnabc},
and equating terms of $O(\gamma)$ we get the equations satisfied by the metric perturbations. 
Defining the dimensionless variables

\be A_1 = (R^9 U_{KK}^3)^{1/2} a_1 , \quad B_1 = (R^9 U_{KK}^3)^{1/2} b_1 , \quad C_1 
= (R^9 U_{KK}^3)^{1/2} c_1,\ee

and changing coordinates to $\eta$, from \C{eqnabc} we get that

\bea \label{moreeqnabc}
&& -8 (1 -\eta^3) C_1'' +\frac{12 C_1'}{\eta} -\frac{5 (1-\eta^3) B_1'}{\eta} +\frac{15 B_1 + 12 
C_1}{\eta^2} \non \\ &=& -\frac{3}{512 \sqrt{\eta}} \Big[ \frac{729}{2} -2106 \eta^3 
+\frac{22221}{2} \eta^6 
- 474174 \eta^9 + \frac{1072869}{2} \eta^{12} \Big] \equiv \frac{f_1 (\eta^3)}{\sqrt{\eta}}, \non \\
&& 12 (1 -\eta^3) C_1'' +\frac{2 (\eta^3 -11) C_1'}{\eta} -\frac{5 (1-\eta^3) A_1'}{6\eta} 
-\frac{5(5 B_1 + 4 C_1)}{\eta^2} \non \\ &=& \frac{1}{1024 \sqrt{\eta}} \Big[ 3753 -19008 \eta^3
+ 105831 \eta^6 - 4273602 
\eta^9 + 4899321 \eta^{12} \Big] \equiv \frac{f_2 (\eta^3)}{\sqrt{\eta}} , \non \eea
\bea
&& -2 (1 -\eta^3) (A_1'' + 9 B_1'' + 3 C_1'') +\frac{3 (7 B_1 -16 C_1)}{\eta^2} 
\non \\ && + \frac{(4 \eta^3 + 5) A_1'}{\eta} 
+\frac{39 (\eta^3 + 1) B_1'}{\eta} 
+\frac{6 (\eta^3 +2) C_1'}{\eta} \non \\
&=& \frac{3}{1024 \sqrt{\eta}} \Big[ 945 + 216 \eta^3
- 20106 \eta^6 + 37581 \eta^9 - 16569 \eta^{12} \Big] \equiv \frac{f_3 (\eta^3)}{\sqrt{\eta}}. \eea

From the equations \C{moreeqnabc}, it follows that $a_1$ has a zero mode given by
$a_1 = {\rm const}$, which is not fixed by the equations of motion. We shall fix its value shortly.

Ignoring the issue of the zero mode of $A_1$ for the time being, we now solve the 
equations \C{moreeqnabc} in order 
to find the 
metric perturbations in the Einstein frame. We find it convenient to further redefine variables

\be \label{multfac}
A_1 = \eta^{3/2} {\mathcal{A}}_1, \quad B_1 = \eta^{3/2} {\mathcal{B}}_1,
\quad C_1 = \eta^{3/2} {\mathcal{C}}_1,\ee

so that the metric perturbations are given by

\be a_1 (U) = \frac{{\mathcal{A}}_1 (U)}{\sqrt{R^9 U^3}} , 
\quad b_1 (U) = \frac{{\mathcal{B}}_1 (U)}{\sqrt{R^9 U^3}} , 
\quad c_1 (U) =\frac{{\mathcal{C}}_1 (U)}{\sqrt{R^9 U^3}} ,\ee

Using this, we see that equations \C{moreeqnabc} reduce to 
  
\bea \label{eqnharmonicabc}
&& -8 (1 -\eta^3) \eta^2 {\mathcal{C}}_1'' +12 (2 \eta^3 -1) \eta {\mathcal{C}}_1' -5 (1-\eta^3) \eta 
{\mathcal{B}}_1' 
\non \\ &&
+ \frac{15 (1 +\eta^3)}{2} {\mathcal{B}}_1 + 6 (\eta^3 + 4) 
{\mathcal{C}}_1 = f_1 (\eta^3) ,  \non \\
&& 12 (1 -\eta^3) \eta^2 {\mathcal{C}}_1'' + 2 (7 - 17 \eta^3) \eta {\mathcal{C}}_1' -\frac{5 (1-\eta^3)}{6} 
\eta {\mathcal{A}}_1' \non \\&&
- \frac{5 (1 -\eta^3)}{4} {\mathcal{A}}_1 -25 {\mathcal{B}}_1 - 2 (22 + 3 \eta^3) 
{\mathcal{C}}_1 = f_2 (\eta^3) ,  \non \\
&& -2 (1 -\eta^3) \eta^2 ( {\mathcal{A}}_1'' +9 {\mathcal{B}}_1'' +3 {\mathcal{C}}_1'' )\non \\&&
+ (10 \eta^3 - 1) \eta {\mathcal{A}}_1' + 3 (31 \eta^3 -5) 
\eta {\mathcal{B}}_1' + 6 (4 \eta^3 -1) \eta {\mathcal{C}}_1' \non \\&&
+ 3(\frac{5 \eta^3}{2} + 2) {\mathcal{A}}_1 +6 (12 \eta^3 + 11) 
{\mathcal{B}}_1 + \frac{3 (9 \eta^3 - 23)}{2} {\mathcal{C}}_1 = f_3 (\eta^3) . \eea

Though the system of equations \C{eqnharmonicabc} looks complicated, it is easy to see that they are solved
by sums of powers of harmonic functions $\eta^3$.  
So we make the ansatz

\be \label{valcoeff}
{\mathcal{A}}_1 (\eta)= \sum_{k=0}^\infty \frac{{\hat{a}}_{3k}}{512} \eta^{3k}, \quad 
{\mathcal{B}}_1 (\eta)= \sum_{k=0}^\infty \frac{{\hat{b}}_{3k}}{512} \eta^{3k}, \quad {\mathcal{C}}_1 
(\eta) =\sum_{k=0}^\infty \frac{{\hat{c}}_{3k}}{512} \eta^{3k},\ee

where ${\hat{a}}_{3k}, {\hat{b}}_{3k}$ and ${\hat{c}}_{3k}$ are numbers. Demanding regularity of the solution
as $U_{KK} \rightarrow 0$, we do not have negative values of $k$ in \C{valcoeff}. Thus including the 
zero mode, $a_1 (U)$ is given by

\be a_1 (U) = {\rm {const}} + \frac{1}{\sqrt{R^9 U^3}} \sum_{k=0}^\infty \frac{{\hat{a}}_{3k}}{512} \Big(
\frac{U_{KK}}{U} \Big)^{3k} . \ee

In the supersymmetric limit $U_{KK} \rightarrow 0$, $a_1 (U)$ is given by

\be a_1 (U) = {\rm {const}} + \frac{{\hat{a}}_0}{512 \sqrt{R^9 U^3}}. \ee

We set the zero mode to zero so that the $x_4$ coordinate is canonically normalized~\cite{Gubser:1998nz}. 
Solving \C{eqnharmonicabc} boils down to solving a system of 
coupled difference equations obtained by equating terms involving the same 
powers of $\eta$. The equations are 

\bea \frac{5 {\hat{b}}_0}{2} + 8 {\hat{c}}_0 &=& -\frac{729}{2}, \non \\
\frac{5 {\hat{a}}_0}{4} + 25 {\hat{b}}_0 + 44 {\hat{c}}_0 &=& -\frac{3753}{2} , \non \\
2 {\hat{a}}_0 + 22 {\hat{b}}_0  -\frac{23 {\hat{c}}_0}{2} &=& \frac{945}{2}, \eea

and

\bea && -4 \Big( k - \frac{1}{2} \Big) \Big( 2 (3 k + 2) {\hat{c}}_{3k} - ( 6 k -5) {\hat{c}}_{3(k-1)} \Big)  
+ \frac{5 (1- 2k)}{2} ({\hat{b}}_{3k} - {\hat{b}}_{3(k-1)} ) \non \\
&&= 2106 \delta_{k,1} - \frac{22221}{2}
\delta_{k,2} + 474174 \delta_{k,3} - \frac{1072869}{2} \delta_{k,4} ,\non \\
&& - \frac{5 ( 2k + 1)}{4} {\hat{a}}_{3k} + \frac{5 ( 2k - 1)}{4} {\hat{a}}_{3(k-1)} 
+ 6 \Big( 18 k^2 + k - \frac{22}{3} \Big) 
{\hat{c}}_{3k} -6 \Big( 18 k^2 - 25 k + 8 \Big) {\hat{c}}_{3(k-1)} \non \\ && - 25 {\hat{b}}_{3k} 
= - 9504 \delta_{k,1}
+ \frac{105831}{2} \delta_{k,2} - 2136801 \delta_{k,3} + \frac{4899321}{2} \delta_{k,4} , \non \\
&& \Big(- 6 k^2 + k + 2 \big) {\hat{a}}_{3k} + \Big( 6 k^2 - 4k + \frac{1}{2} \Big) 
{\hat{a}}_{3(k-1)} + \Big( - 54 k^2 + 3 k + 22 \Big) {\hat{b}}_{3k} \non \\ &&+ 3 
\Big( 18 k^2 - 11 k + 1 \Big) 
{\hat{b}}_{3(k-1)} -
\Big( 18 k^2 + \frac{23}{2} \Big)  {\hat{c}}_{3k} + 9 \Big( 2 k^2 - 2 k 
+ \frac{1}{2} \Big) {\hat{c}}_{3(k-1)} \non \\ &&  
=  108 \delta_{k,1} - 10053 \delta_{k,2} + \frac{37581}{2} \delta_{k,3} - \frac{16569}{2} 
\delta_{k,4} ,\eea

for $k \geq 1$. These equations can be solved recursively and yield the solution

\bea \label{valval}
{\mathcal{A}}_1 (\eta) &=& \frac{1}{512} \Big[ \frac{2034}{5} + \frac{41106}{5} \eta^3 -\frac{20491152}{175} 
\eta^6 + \frac{5360956}{21} \eta^9 - \frac{1116115426}{17325} \eta^{12} 
+ \ldots \Big]  , \non \\
{\mathcal{B}}_1 (\eta) &=& \frac{1}{512} \Big[ -\frac{169}{5} - 845 \eta^3 + \frac{454014}{35} \eta^6 
- \frac{129590014}{4725} \eta^9 + \frac{391267183}{51975} \eta^{12} + \ldots\Big]  , \non \\
{\mathcal{C}}_1 (\eta) &=& \frac{1}{512} \Big[ -35 - \frac{37}{5} \eta^3 - \frac{33843}{35} \eta^6 
- \frac{81526}{189} \eta^9 - \frac{719803}{1485} \eta^{12} + \ldots\Big]  ,\eea

where $\ldots$ are the terms higher order in $\eta$, starting from $\eta^{15}$. We do not have a closed form
expression for the metric perturbations. Thus, for example, we cannot determine 
the precise 
nature of the perturbations near $U = U_{KK}$. Note that the metric perturbations vanish 
as $U \rightarrow \infty$,
i.e., far away from the color branes.  

\section{Obtaining the perturbed dilaton}

We next calculate the dilaton perturbation at $O(\gamma)$, for which we need to 
consider the action \C{valmetric} 
at $O(\gamma^2)$. Including only the terms that depend on the dilaton, this gives

\be S_{\phi}^{\gamma^2} = S_{\phi_2}^{\gamma^2} + S_{\phi_1}^{\gamma^2}, \ee 

where

\be \label{phi2zero}
S_{\phi_2}^{\gamma^2} = V  \int_{U_{KK}}^\infty dU \sqrt{\tilde{\hat{g}}} \Big[ -\frac{3\p_U
\phi_2}{4U H^2 P^2} - \frac{g_s^2 Q^2}{4 L^8} \Big( \frac{U}{R} \Big)^{3/8} \phi_2 \Big], \ee

which depends on $\phi_2$, and

\bea \label{phi21}
S_{\phi_1}^{\gamma^2} = V  \int_{U_{KK}}^\infty dU \sqrt{\tilde{\hat{g}}} \Big[ -\frac{3\p_U
\phi_1}{4U H^2 P^2} -  \frac{(\p_U \phi_1)^2}{2 H^2 P^2}
- \frac{g_s^2}{4 \cdot 4!} \Big( \frac{U}{R} \Big)^{3/8} \Big( \phi_1 + \frac{\phi_1^2}{4} \Big)
{\hat{F}}_4^2 \non \\
- \frac{3}{2} \gamma \Big( \frac{U}{R} \Big)^{-9/8}  \hat{W} \phi_1\Big],  \eea

which depends on $\phi_1$.
In \C{phi2zero}, $\phi_2 \sim O(\gamma^2)$, and so we substitute the supergravity values of 
the metric and the four 
form flux. Thus $S_{\phi_2}^{\gamma^2}$ vanishes 
for the same reason as in \C{phione}. In \C{phi21}, we substitute the
supergravity values of the various fields in the terms involving $\phi_1^2$ and 
in the term involving $\hat{W} 
\phi_1$, 
while in the remaining terms we substitute the values of the fields at $O(\gamma)$.    
Now $\hat{W}$ for the supergravity metric \C{metEin} is given by

\be \label{hatback}
\hat{W} = \frac{9\Big( 135 U^{12} + 819U^6 U_{KK}^6 -756 U^3 U_{KK}^9
+ 4739 U_{KK}^{12} \Big)}{1024 U^{25/2} R^{15/2}}. \ee

Note that \C{hatback} does not vanish for $U_{KK} =0$, as the theory is not conformal.
Thus defining $\phi_1 = \gamma {\hat\phi}_1$, we get that

\bea 
S_{\phi_1}^{\gamma^2} &=& \gamma^2 V  \int_{U_{KK}}^\infty dU \Big[ - \frac{f (U) U^4 (\p_U {\hat\phi}_1 
)^2}{2} - \frac{9 U^2 {\hat\phi}_1^2}{16} \non \\ &&- \frac{3 f(U) U^3}{4} \Big( a_1 + 8 b_1 + 4 c_1 
\Big) (\p_U {\hat\phi}_1) 
-\frac{9 U^2}{4} \Big( a_1 + 10 b_1 -4 c_1 \Big) {\hat\phi}_1 \non \\ &&-\frac{27 \Big(
135 U^{12} + 819U^6 U_{KK}^6 -756 U^3 U_{KK}^9
+ 4739 U_{KK}^{12} \Big) {\hat\phi}_1}{2048 U^{23/2} R^{9/2}}
\Big].\eea

This leads to the equation of motion 

\bea \label{eqnphi} &&(1 - \eta^3) \varphi_1'' - \frac{(\eta^3 + 2)}{\eta} \varphi_1' - \frac{9 
\varphi_1}{8 \eta^2} \non \\ &=& \frac{9}{4 \sqrt{\eta}} \Big(
{\mathcal{A}}_1 + 10 {\mathcal{B}}_1 -4 {\mathcal{C}}_1 \Big)
- \frac{9 (\eta^3 + 1)}{8 \sqrt{\eta}} \big({\mathcal{A}}_1 + 8 {\mathcal{B}}_1 +4 {\mathcal{C}}_1 \Big) 
\non \eea
\bea
&& + \frac{3 (1 - \eta^3) \sqrt{\eta}}{4} 
\big({\mathcal{A}}_1' + 8 {\mathcal{B}}_1' +4 {\mathcal{C}}_1' \Big)  \non \\&&+ 
\frac{27}{2048 \sqrt{\eta}} \Big( 135  + 819 \eta^6 -756 \eta^9 + 4739 \eta^{12} \Big) ,\eea

where $\varphi_1$ is the dimensionless dilaton perturbation defined by

\be \varphi_1 =  (R^9 U_{KK}^3)^{1/2} {\hat\phi}_1, \ee

and all derivatives in \C{eqnphi} are with respect to $\eta$. Just like the
metric perturbations, defining $\varphi_1 = \eta^{3/2} 
{\hat\varphi}_1$, \C{eqnphi} leads to

\bea  \label{dilpert}
&&(1 - \eta^3) \eta^2 {\hat\varphi}_1'' + (1 - 4 \eta^3) \eta {\hat\varphi}_1' - \frac{9 (\eta^3 + 
\frac{3}{2})}{4} {\hat\varphi}_1 \non \\
&=& \frac{9}{4} \Big(
{\mathcal{A}}_1 + 10 {\mathcal{B}}_1 -4 {\mathcal{C}}_1 \Big)
- \frac{9 (\eta^3 + 1)}{8} \big({\mathcal{A}}_1 + 8 {\mathcal{B}}_1 +4 {\mathcal{C}}_1 \Big) 
\non \\
&&+ \frac{3 (1 - \eta^3) \eta}{4} 
\big({\mathcal{A}}_1' + 8 {\mathcal{B}}_1' +4 {\mathcal{C}}_1' \Big)  \non \\ &&+ 
\frac{27}{2048} \Big( 135  + 819 \eta^6 -756 \eta^9 + 4739 \eta^{12} \Big).\eea

which we solve by making the ansatz

\be {\hat\varphi}_1 (\eta) = \sum_{k =0}^\infty \frac{{\tilde\varphi}_{3k}}{512} \eta^{3k}.\ee

The difference equations are 

\be{\tilde\varphi}_0 = - \frac{{\hat{a}}_0}{3} -4 {\hat{b}}_0 + 4 {\hat{c}}_0  - 270 ,\ee

and

\bea && \Big( k^2 -  \frac{3}{8} \Big)  {\tilde\varphi}_{3k} - \Big( k^2 - k 
+ \frac{1}{4} \Big) {\tilde\varphi}_{3(k-1)} \non \\
&=&   \frac{k + \frac{1}{2}}{4} {\hat{a}}_{3k} - \frac{k - \frac{1}{2}}{4} {\hat{a}}_{3(k-1)} 
 + (2 k + \frac{3}{2}) {\hat{b}}_{3k} - (2k -1){\hat{b}}_{3(k-1)} \non \\&&+ \Big(
k - \frac{3}{2} \Big)
{\hat{c}}_{3k} -\Big( k -\frac{1}{2} \Big) {\hat{c}}_{3(k-1)}  
+ \frac{3}{4} \Big( 819 \delta_{k,2} 
-756 \delta_{k,3} + 4739 \delta_{k,4} \Big),  \eea

for $k \geq 1$. Thus on using \C{valval}, this leads to

\be \label{valdil}
{\hat\varphi}_1 (\eta) = \frac{1}{512} \Big[ - \frac{2052}{5} 
+ \frac{216}{5} \eta^3 - \frac{17172}{29}
\eta^6 + \frac{9086952}{3335} \eta^9 + \frac{130277364}{416875} \eta^{12} 
+ \ldots \Big]  \ee

where $\ldots$ are terms of $O(\eta^{15})$.






\providecommand{\href}[2]{#2}\begingroup\raggedright\endgroup

\end{document}